\documentclass[12pt]{article}
\usepackage{axodraw,bbold}

\catcode`@=12
\begin{document}
\thispagestyle{empty}
\begin{flushright} 
UCRHEP-T402\\ 
November 2005\
\end{flushright}
\vspace{0.5in}
\begin{center}
{\LARGE	\bf Tribimaximal neutrino mixing\\
from a supersymmetric model\\
with $A_4$ family symmetry\\}
\vspace{1.5in}
{\bf Ernest Ma\\}
\vspace{0.2in}
{\sl Physics Department, University of California, Riverside, 
California 92521\\}
\vspace{1.5in}
\end{center}

\begin{abstract}\
In the supersymmetric seesaw model of neutrino masses, augmented by the 
non-Abelian discrete tetrahedral symmetry $A_4$, a specific pattern of 
neutrino mixing is \underline{automatically} \underline{generated} if one 
of the three heavy singlet neutrino superfields acquires a nonzero vacuum 
expectation value. This pattern turns out to be exactly that of tribimaximal 
mixing, i.e. $\sin^2 \theta_{23} = 1/2$, $\sin^2 \theta_{12} = 1/3$, and 
$\sin^2 \theta_{13} = 0$, in good agreement with data.
\end{abstract}

\newpage
\baselineskip 24pt

In the well-known canonical seesaw mechanism \cite{seesaw}, three heavy 
singlet Majorana neutrinos $N_i$ ($i=1,2,3$) are added to the Standard 
Model of elementary particles, so that
\begin{equation}
{\cal M}_\nu^{(e,\mu,\tau)} = -{\cal M}_D {\cal M}_N^{-1} {\cal M}_D^T,
\end{equation}
where ${\cal M}_D$ is the $3 \times 3$ Dirac mass matrix linking the 
observed neutrinos $\nu_\alpha$ ($\alpha=e,\mu,\tau$) to $N_i$, and 
${\cal M}_N$ is the Majorana mass matrix of $N_i$.  Consider now its 
diagonalization, i.e.
\begin{equation}
{\cal M}_\nu^{(e,\mu,\tau)} = U_{\alpha i} \pmatrix{m_1 & 0 & 0 \cr 0 & m_2 
& 0 \cr 0 & 0 & m_3} U^T_{j \beta}.
\end{equation}
Present neutrino-oscillation data have determined the absolute values of 
$U_{\alpha i}$ to a large extent, as well as the two differences of the 
absolute squares of the three masses \cite{lp05}.  Theoretically, the 
obvious challenge is to find a simple and natural understanding of these 
results.

In the following, it will be shown that in the context of supersymmetry, 
augmented by the non-Abelian discrete symmetry $A_4$ \cite{mr01}, a 
specific three-parameter form of ${\cal M}_\nu^{(e,\mu,\tau)}$ 
is \underline{automatically} \underline{generated} if one of the three $N_i$ 
superfields acquires a nonzero vacuum expectation value.  This results in 
a specific $U_{\alpha i}$ which turns out to be exactly that of the so-called 
tribimaximal mixing of Harrison, Perkins, and Scott \cite{hps02}, i.e.
\begin{equation}
U_{\alpha i} = \pmatrix{ \sqrt{2/3} & \sqrt{1/3} & 0 \cr -\sqrt{1/6} & 
\sqrt{1/3} & -\sqrt{1/2} \cr -\sqrt{1/6} & \sqrt{1/3} & \sqrt{1/2}}.
\end{equation}
In terms of the usual neutrino-oscillation parameters, this means that
\begin{equation}
\sin^2 \theta_{23} = {1 \over 2}, ~~~ 
\sin^2 \theta_{12} = {1 \over 3}, ~~~ 
\sin^2 \theta_{13} = 0,
\end{equation}
in good agreement with data \cite{lp05}.

The non-Abelian finite group $A_4$ is the symmetry group of the even 
permutation of four objects.  It is also the symmetry group of the regular 
tetrahedron, one of five perfect geometric solids which was identified by 
Plato with the Greek element ``fire'' \cite{plato}.  There are twelve group 
elements and four irreducible representations: \underline{1}, 
\underline{1}$'$, \underline{1}$''$, and \underline{3}.  Let $a_{1,2,3}$ 
and $b_{1,2,3}$ transform as \underline{3} under $A_4$, then \cite{fuji}
\begin{eqnarray}
&& a_1 b_1 + a_2 b_2 + a_3 b_3 \sim \underline{1}, \\ 
&& a_1 b_1 + \omega^2 a_2 b_2 + \omega a_3 b_3 \sim \underline{1}', \\ 
&& a_1 b_1 + \omega a_2 b_2 + \omega^2 a_3 b_3 \sim \underline{1}'', \\ 
&& (a_2 b_3, a_3 b_1, a_1 b_2) \sim \underline{3}, \\ 
&& (a_3 b_2, a_1 b_3, a_2 b_1) \sim \underline{3}, 
\end{eqnarray}
where $\omega = \exp(2 \pi i/3) = -1/2 + i \sqrt{3}/2$.

Under $A_4$, the lepton doublets $(\nu_i,l_i)$ transform as \underline{3} and 
the charged-lepton singlets $l^c_i$ as \underline{1}, \underline{1}$'$, 
\underline{1}$''$, with three Higgs doublets $(\phi^0_i, \phi^-_i)$ 
transforming as \underline{3}.  Assuming equal $\langle \phi^0_i \rangle = 
v$, the charged-lepton mass matrix linking $l_i$ to $l^c_j$ is then given 
by \cite{mr01}
\begin{equation}
{\cal M}_l = U_L \pmatrix{h_e & 0 & 0 \cr 0 & h_\mu & 0 \cr 0 & 0 & 
h_\tau} \sqrt{3} v,
\end{equation}
where
\begin{equation}
U_L = {1 \over \sqrt 3} \pmatrix{1 & 1 & 1 \cr 1 & \omega & \omega^2 \cr 
1 & \omega^2 & \omega}.
\end{equation}

In the neutrino sector, the three singlets $N_i$ transform as \underline{3} 
under $A_4$ with one Higgs doublet $(\eta^+,\eta^0)$ transforming as 
\underline{1}.  Hence
\begin{equation}
{\cal M}_D = U_L^\dagger \pmatrix{m_D & 0 & 0 \cr 0 & m_D & 0 \cr 0 & 0 & m_D},
\end{equation}
and
\begin{equation} 
{\cal M}_{N} = \pmatrix{M & 0 & 0 \cr 0 & M & 0 \cr 0 & 0 & M}. 
\end{equation}
The resulting ${\cal M}_\nu$ in the $(e, \mu, \tau)$ basis is then given by
\begin{equation}
{\cal M}_\nu = U_L^\dagger \pmatrix{m_0 & 0 & 0 \cr 0 & m_0 & 0 \cr 0 & 0 & 
m_0} U_L^* = \pmatrix{m_0 & 0 & 0 \cr 0 & 0 & m_0 \cr 0 & m_0 & 0},
\end{equation}
where $m_0 = -m_D^2/M$.  This is the starting point of the two original $A_4$ 
models \cite{mr01,bmv03}.  Since all three neutrinos have the same absolute 
mass, there is actually no mixing in this case.  Whereas small radiative 
perturbations can result in a realistic mass matrix 
\cite{plato,bmv03,hrsvv04}, $U_{\alpha i}$ is not completely predicted 
in this approach.

Here it is proposed that ${\cal M}_N$ is actually of the form
\begin{equation}
{\cal M}_N = \pmatrix{A & 0 & 0 \cr 0 & B & C \cr 0 & C & B}.
\end{equation}
The justification of this in terms of the superpotential of $N_i$ 
will be discussed in detail later.  For now, just consider the resulting 
$3 \times 3$ Majorana neutrino mass matrix in the $(e, \mu, \tau)$ basis. 
Using Eqs.~(1), (11), (12), and (15), ${\cal M}_\nu$ is then given by
\begin{equation}
{-m_D^2 \over 3A(B^2-C^2)} \pmatrix{B^2-C^2+2AB-2AC & B^2-C^2-AB+AC & 
B^2-C^2-AB+AC \cr B^2-C^2-AB+AC & B^2-C^2-AB-2AC & B^2-C^2+2AB+AC \cr 
B^2-C^2-AB+AC & B^2-C^2+2AB+AC & B^2-C^2-AB-2AC}.
\end{equation}
This matrix is a special form of the four-parameter matrix proposed in 
Ref.~\cite{m04}, i.e.
\begin{equation}
{\cal M}_\nu = \pmatrix{a+(2d/3) & b-(d/3) & c-(d/3) \cr b-(d/3) & c+(2d/3) & 
a-(d/3) \cr c-(d/3) & a-(d/3) & b+(2d/3)},
\end{equation}
with
\begin{equation}
a = {-m_D^2 (B^2-C^2+2AB) \over 3A(B^2-C^2)}, ~~~ 
b = c  = {-m_D^2 (B^2-C^2-AB) \over 3A(B^2-C^2)}, ~~~ 
d = {m_D^2 C \over B^2-C^2}.
\end{equation}
As promised, it is exactly diagonalized by Eq.~(3), i.e.
\begin{equation}
{\cal M}_\nu = \pmatrix{ \sqrt{2/3} & \sqrt{1/3} & 0 \cr -\sqrt{1/6} & 
\sqrt{1/3} & -\sqrt{1/2} \cr -\sqrt{1/6} & \sqrt{1/3} & \sqrt{1/2}}
\pmatrix{m_1 & 0 & 0 \cr 0 & m_2 & 0 \cr 0 & 0 & m_3}  
\pmatrix{ \sqrt{2/3} & -\sqrt{1/6} & -\sqrt{1/6} \cr \sqrt{1/3} & 
\sqrt{1/3} & \sqrt{1/3} \cr 0 & -\sqrt{1/2} & \sqrt{1/2}},
\end{equation}
with
\begin{equation}
m_1 = {-m_D^2 \over B+C}, ~~~ 
m_2 = {-m_D^2 \over A}, ~~~ 
m_3 = {m_D^2 \over B-C}.
\end{equation}
Since there are three independent parameters $(A,B,C)$, it is clear that 
the three neutrino masses may be chosen arbitrarily to fit the data. 
In other words, this model predicts $U_{\alpha i}$ but not $m_{1,2,3}$.

To obtain ${\cal M}_N$ of Eq.~(15), consider the most general superpotential 
of $N_i$ invariant under $A_4$ up to quartic terms, i.e.
\begin{eqnarray}
W &=& {1 \over 2} m_N (N_1^2 + N_2^2 + N_3^2) + f N_1 N_2 N_3 \nonumber \\ 
&+& {\lambda_1 \over 4 M_{Pl}} (N_1^4 + N_2^4 + N_3^4) + {\lambda_2 \over 
2 M_{Pl}} (N_2^2 N_3^2 + N_3^2 N_1^2 + N_1^2 N_2^2),
\end{eqnarray}
where $M_{Pl} = 1.2 \times 10^{19}$ GeV is the Planck mass.  To preserve 
the supersymmetry of the complete theory at this high scale, a solution 
must exist for which the minimum of the resulting scalar potential
\begin{eqnarray}
V &=& |m_N N_1 + f N_2 N_3 + {\lambda_1 \over M_{Pl}} N_1^3 + 
{\lambda_2 \over M_{Pl}} N_1 (N_2^2+N_3^2)|^2 \nonumber \\
&+& |m_N N_2 + f N_3 N_1 + {\lambda_1 \over M_{Pl}} N_2^3 + 
{\lambda_2 \over M_{Pl}} N_2 (N_3^2+N_1^2)|^2 \nonumber \\ 
&+& |m_N N_3 + f N_1 N_2 + {\lambda_1 \over M_{Pl}} N_3^3 + 
{\lambda_2 \over M_{Pl}} N_3 (N_1^2+N_2^2)|^2 
\end{eqnarray}
is zero.  The only solution which has ever been assumed up to now is 
$\langle N_{1,2,3} \rangle = 0$, for which ${\cal M}_N$ is indeed given by 
Eq.~(13).  However, there is another natural solution, i.e.
\begin{equation}
\langle N_{2,3} \rangle = 0, ~~~ \langle N_1 \rangle^2 = {-m_N M_{Pl} \over 
\lambda_1}.
\end{equation}
In that case, the mass term corresponding to the shifted field $N'_1 \equiv 
N_1 - \langle N_1 \rangle$ in $W$ becomes
\begin{equation}
m_N + {3 \lambda_1 \langle N_1 \rangle^2 \over M_{Pl}} = -2m_N,
\end{equation}
and $N_2 N_3$ has the mass term $f \langle N_1 \rangle$, whereas $N_2^2$ and 
$N_3^2$ have the mass term
\begin{equation}
m_N + {\lambda_2 \langle N_1 \rangle^2 \over M_{Pl}} = m_N \left( 1 - {\lambda_2 \over \lambda_1} \right).
\end{equation}
In other words, Eq.~(15) is 
\underline{automatically} \underline{generated} with $A=-2m_N$, 
$B=(1-\lambda_2/\lambda_1)m_N$, and $C=f\langle N_1 \rangle$ which 
is of order $A$ and $B$ if $f$ is of order $|\lambda_1 m_N/M_{Pl}|^{1/2}$.

Since the superpotential also contains the term
\begin{equation}
h[(\nu_1 N_1 + \nu_2 N_2 + \nu_3 N_3)\eta^0 - 
(l_1 N_1 + l_2 N_2 + l_3 N_3)\eta^+],
\end{equation}
the soft term
\begin{equation}
-h\langle N_1 \rangle (\nu_1 \eta^0 - l_1 \eta^+)
\end{equation}
must be added to allow $(\nu_1,l_1)$ and $(\eta^+,\eta^0)$ to remain 
massless at this high scale.  This is of course fine tuning, but once it 
is done, it is protected by the exact $R$-parity and supersymmetry of the 
residual theory.  It is analogous to the usual situation in the Minimal 
Supersymmetric Standard Model, where the term $(\nu_i \eta^0 - l_i \eta^+)$ 
is allowed by all its gauge symmetries, but simply forbidden by the 
imposition of $R$-parity, i.e. whatever the allowed term is, a term is 
added to cancel it exactly.

It should be noted that the symmetry being broken at the large scale is 
$A_4$.  Because of the explicit trilinear term $N_1 N_2 N_3$ in the 
superpotential, there is no additional discrete symmetry involved in the 
breaking.  In other words, the concept of $R$-parity does not appear at 
this point.  Below the breaking scale, with the addition of the 
above-mentioned soft term, the concept of $R$-parity emerges for the 
first time, and applies only to the superfields of the Minimal Supersymmetric 
Standard Model.  It does not apply to the $N$ superfields because they have 
all been integrated away.  This is perfectly consistent with an effective 
supersymmetric field theory at the electroweak scale with Majorana neutrino 
masses.

If soft terms which break $A_4$ are simply added to the Majorana mass matrix 
of $N_i$, the same model below the seesaw scale can be obtained.  There are 
however two important differences.  One is that whereas this procedure 
may be used to obtain any pattern that is desired, the procedure advocated 
here will only result in the particular pattern shown.  The other is that 
the two models have different interactions above the seesaw scale.  Even 
though they are experimentally indistinguishable at present energies, they 
are at least theoretically distinct.

To avoid having three Higgs doublet superfields $(\phi^0_i,\phi^-_i)$ and 
their three partners at the electroweak scale, this model can be modified by 
having just one $(\phi^0,\phi^-) \sim \underline{1}$ under $A_4$, but with 
the addition of three heavy singlets $\zeta_i$ which transform as 
\underline{3} under $A_4$.  The Yukawa coupling terms in the charged-lepton 
sector are then given by \cite{fn79,hiding}
\begin{equation}
{h_{ijk} \over \Lambda} (\nu_i \phi^- - l_i \phi^0) l^c_j \zeta_k.
\end{equation}
To decouple $\zeta_i$ from $N_i$, an extra $Z_4$ symmetry is assumed, under 
which the only nontrivial transformations are $\zeta \sim i$ and 
$l^c \sim -i$.  Consider then the superpotential
\begin{eqnarray}
W_\zeta = {1 \over 2} m_\zeta (\zeta_1^2 + \zeta_2^2 + \zeta_3^2) + 
{\lambda_3 \over 4 M_{Pl}} (\zeta_1^4 + \zeta_2^4 + \zeta_3^4)  
+ {\lambda_4 \over 2 M_{Pl}} (\zeta_2^2 \zeta_3^2 + \zeta_3^2 \zeta_1^2 + 
\zeta_1^2 \zeta_2^2),
\end{eqnarray}
where $m_\zeta$ breaks $Z_4$ softly.  The resulting scalar potential is
\begin{eqnarray}
V_\zeta &=& | m_\zeta \zeta_1 + {\lambda_3 \over M_{Pl}} \zeta_1^3 + 
{\lambda_4 \over M_{Pl}} \zeta_1 (\zeta_2^2 + \zeta_3^2) |^2 \nonumber \\
&+& | m_\zeta \zeta_2 + {\lambda_3 \over M_{Pl}} \zeta_2^3 + 
{\lambda_4 \over M_{Pl}} \zeta_2 (\zeta_3^2 + \zeta_1^2) |^2 \nonumber \\
&+& | m_\zeta \zeta_3 + {\lambda_3 \over M_{Pl}} \zeta_3^3 + 
{\lambda_4 \over M_{Pl}} \zeta_3 (\zeta_1^2 + \zeta_2^2) |^2,
\end{eqnarray}
which has the desired solution
\begin{equation}
\langle \zeta_1 \rangle = \langle \zeta_2 \rangle = \langle \zeta_3 \rangle = 
{-m_\zeta M_{Pl} \over \lambda_3 + 2 \lambda_4},
\end{equation}
for which the supersymmetry is unbroken.

Other realizations of Eq.~(3) also exist \cite{af05,m05,bh05,gl05}.  They can 
be classified according to the four parameters $(a,b,c,d)$ of Eq.~(17) as 
follows.  In Ref.~\cite{af05}, it is 
proposed that $b=c=0$.  In Ref.~\cite{m05}, the case $a=0$ and $b=c$ is 
discussed.  In Ref.~\cite{bh05}, the conditions are $b=c$ and  $d^2 = 
3b(b-a)$.  Here and in Ref.~\cite{gl05}, Eq.~(17) is reduced by only $b=c$.  
(All these examples are based on $A_4$ except the last one, which is based 
on the Coxeter group $B_4$, which is also the symmetry group of the 
hyperoctahedron \cite{b4}.)  What sets the present model apart from all 
others is the automatic generation of Eq.~(15), using 
the hitherto unrecognized possibility of Eq.~(23).

In conclusion, it has been shown how the tribimaximal mixing pattern of 
neutrinos can be derived in the supersymmetric seesaw model with $A_4$ 
symmetry.  The spontaneous breaking of $A_4$ through the nonzero vacuum 
expectation value of one of the three heavy singlet neutrino superfields 
\underline{automatically} \underline{generates} the desired neutrino 
mass matrix.  Below the seesaw scale, the model is identical to that of 
the Minimal Supersymmetric Standard Model, but with arbitrary 
nonzero Majorana neutrino masses which mix tribimaximally \cite{lx05}.

This work is supported in part by the U.~S,~Department of Energy under Grant 
No.~DE-FG03-94ER40837.

\bibliographystyle{unsrt}

\end{document}